\documentclass[10pt,journal,cspaper,compsoc,twocolumn ]{IEEEtran}
\usepackage{url}
\usepackage{amsmath}
\usepackage{amsfonts}
 \usepackage{theorem}
\usepackage[dvips]{epsfig}
\usepackage{graphicx}
\usepackage{enumerate}

 \usepackage{verbatim}
 \usepackage{amssymb}
\usepackage{amsbsy}
 \usepackage{setspace}
\usepackage{url}

\theoremstyle{plain}
\newtheorem{Theorem}{Theorem}

\newtheorem{Proposition}{Proposition}
\newtheorem{Corollary}{Corollary}

\newtheorem{Problem}{Problem}
\newtheorem{Fact}{Fact}

{\theorembodyfont{\rmfamily} \newtheorem{Remark}{Remark}}
{\theorembodyfont{\rmfamily} }
{\theorembodyfont{\rmfamily} \newtheorem{Example}{Example}}
{\theorembodyfont{\rmfamily} }
{\theorembodyfont{\rmfamily} }

\newcommand {\R}{\mathbb R}

\newcommand{\be}{\begin{equation}}
\newcommand{\ee}{\end{equation}}

\newcommand{\Int}{\operatorname{{\mathrm int}}}

\newcommand{\LMD}{\lambda_0,\dots,\lambda_n}

\newcommand{\LMDSTAR}{\lambda_0^*,\dots,\lambda_n^*}

\usepackage{datetime}

\begin{document}

\title{On the Structure of the Initiation and Elongation Rates
that Maximize Protein Production  in the Ribosome Flow Model}
\author{Yoram Zarai and  Michael Margaliot\thanks{The authors are with the School of Elec. Eng.-Systems, Tel Aviv University, Israel.  Corresponding author: Prof. Michael Margaliot,
School of Elec. Eng.-Systems, Tel Aviv University, Israel 69978. Tel +972-3-6407768. Email
\texttt{michaelm@eng.tau.ac.il} }}

\maketitle

\begin{abstract}
Translation is a crucial  step in gene expression.
During translation,  macromolecules called ribosomes
``read'' the mRNA strand in a sequential manner
and produce a corresponding protein.
Translation is known to consume most of the cell's energy.
Maximizing  the protein production rate in mRNA translation,
 subject to  the  bounded  biomolecullar  budget, is thus an important problem in both biology
 and biotechnology.

We consider  this problem using
a mathematical model for   mRNA
translation called the ribosome flow model~(RFM).
For an mRNA strand with~$n$ sites the RFM includes
 $n$ state-variables that encode the normalized ribosomal  density
 at each site, and~$n+1$ positive parameters: the initiation  rate and elongation rates
 along the chain.
 An affine constraint on these rates is used to model
 the bounded cellular budget.

We show that for a homogeneous constraint
the rates that maximize the steady-state protein production rate
 have a special structure. They are symmetric with respect to
the middle of the chain, and monotonically increase as we move towards the center of the chain.
The  ribosomal densities corresponding to the optimal rates monotonically
decrease along the chain. We discuss some of the biological implications of these results.
 \end{abstract}


\section{Introduction}
The process in which the genetic information stored in the DNA is transformed into functional proteins is called \emph{gene expression}. Two important
 steps in gene expression are   \emph{transcription} of the DNA code into messenger RNA (mRNA) by RNA polymerase, and then \emph{translation} of the mRNA  into proteins. During   translation,  macro-molecules called ribosomes traverse the mRNA strand, decoding it codon by codon into a corresponding chain of amino-acids which is then folded to become a functional protein. The rate in which proteins are produced during the translation step is called the protein production rate, or the translation rate.

Translation occurs in all organisms and under almost all conditions. Understanding the various factors that affect  this dynamical
process   has important implications to many scientific disciplines, including medicine, evolutionary biology, synthetic biology, and more.

Computational models of translation   are becoming increasingly important as the amount of experimental findings  related to translation increases rapidly~(see, e.g.,
   \cite{Zhang1994,Dana2011,Heinrich1980,MacDonald1968,TullerGB2011,Tuller2007,Chu2012,Shah2013,Deneke2013,Racle2013}). Such models are particulary important in the context
   of synthetic biology and biotechnology, as they
can provide predictions on the qualitative and quantitative
effects of  various manipulations of the
genetic machinery on  the protein production rate.

Translation    is known to consume most
of the cell's energy~\cite{Plotkin2011,Tuller2010,Alberts2002}. Thus,
it is natural to expect that evolution shaped this process
so that it maximizes the protein production rate, given the limited biomolecular budget.
Optimizing  the translation rate is also important  in
  biotechnology where an important objective is to maximize
 the translation efficiency and protein levels of heterologous genes in a new host (see, e.g.,~\cite[Chapter~9]{gene_cloning_book}.

A standard model for translation-elongation is the \emph{totally asymmetric simple exclusion process} (TASEP)~\cite{Shaw2003,TASEP_tutorial_2011}.
 In this model particles stochastically hop along an ordered lattice of  sites.
Simple exclusion means that a particle cannot hop into a site that
is occupied by another particle. This models hard  exclusion between the particles.
In the context of translation, the
 lattice is the mRNA strand; the
  particles are the ribosomes; and hard  exclusion means that a ribosome cannot
  move forward if the codon in front of it is covered by another ribosome.
 TASEP is a
 fundamental model in non-equilibrium statistical mechanics that has been used to model numerous natural and artificial  processes
 including traffic flow, surface growth, communication networks
 and more~\cite{TASEP_book,tasep_ad_hoc_nets}.

The \emph{ribosome flow model}~(RFM)~\cite{reuveni} is the dynamic mean-field approximation of TASEP (see, e.g.,~\cite[section 4.9.7]{TASEP_book} and \cite[p. R345]{solvers_guide}). In the RFM, mRNA molecules are coarse-grained into $n$ consecutive sites.
 The model includes~$n$ non-linear first-order ordinary differential equations:
\begin{align}\label{eq:rfm}
                    \dot{x}_1&=\lambda_{0} (1-x_1) -\lambda_1 x_1(1-x_2), \nonumber \\
                    \dot{x}_2&=\lambda_{1} x_{1} (1-x_{2}) -\lambda_{2} x_{2} (1-x_3) , \nonumber \\
                    \dot{x}_3&=\lambda_{2} x_{ 2} (1-x_{3}) -\lambda_{3} x_{3} (1-x_4) , \nonumber \\
                             &\vdots \nonumber \\
                    \dot{x}_{n-1}&=\lambda_{n-2} x_{n-2} (1-x_{n-1}) -\lambda_{n-1} x_{n-1} (1-x_n), \nonumber \\
                    \dot{x}_n&=\lambda_{n-1}x_{n-1} (1-x_n) -\lambda_n
x_n.
\end{align}
The state variables $x_i(t): \R_+ \to [0,1]$, $i=1,\dots,n$, describe the occupancy level of site $i$ at time $t$, where $x_i(t)=1$ [$x_i(t)=0$] indicates that site $i$ is completely full [empty] at time $t$. The model includes~$n+1$ positive parameters that control the transition rate between the sites: the initiation rate into the chain,
 denoted~$\lambda_0$, and the
 elongation (or transition) rate  between site~$i$ and site~$i+1$,
 denoted~$\lambda_i$, $i=1,\dots,n$.

The rate of ribosome flow from site~$i$ to site~$i+1$ is~$\lambda_{i} x_{i}(t)
(1 - x_{i+1}(t) )$. This rate increases with~$x_i(t)$ (i.e., when site~$i$ is fuller)
and decreases with~$x_{i+1}(t)$ (i.e., when  the consecutive site is becoming fuller).
 In particular, when~$x_{i+1}(t)=1$ (i.e., site~$i+1$ is completely full)
 the rate decreases to zero.
 This may be interpreted as  ``soft exclusion''.
 The term $R(t):=\lambda_n x_n(t)$ describes the rate of ribosomes exiting the mRNA chain, so~$R(t) $ is the protein production  rate at time~$t$.

If we define $x_0(t):=1$ and $x_{n+1}(t):=0$
then~\eqref{eq:rfm} can be written more succinctly as
\be\label{eq:rfm_all}
\dot{x}_i=\lambda_{i-1}x_{i-1}(1-x_i)-\lambda_i x_i(1-x_{i+1}),\quad i=1,\dots,n.
\ee

Let~$x(t,a)$ denote the solution of~\eqref{eq:rfm}
at time~$t \ge 0$ for the initial
condition~$x(0)=a$. Since the  state-variables correspond to normalized occupation levels,
  we always assume that~$a$ belongs to the  closed $n$-dimensional
  unit cube:
$
           C^n:=\{x \in \R^n: x_i \in [0,1] , i=1,\dots,n\}.
$
It is straightforward to verify that this implies that~$x(t,a) \in C^n$ for all~$t\geq0$.
In other words,~$C^n$ is an invariant set of the dynamics~\cite{RFM_stability}.

Let~$\Int(C^n)$ denote the interior of~$C^n$.
It was shown in~\cite{RFM_stability} that the RFM is a
\emph{monotone dynamical system}~\cite{hlsmith}
and that this implies that~\eqref{eq:rfm}
admits a \emph{unique} steady-state point~$e \in \Int(C^n)$.
For~$x=e$ the left-hand side of all the equations
in~\eqref{eq:rfm} is zero, so
\begin{align} \label{eq:ep}
                      \lambda_0 (1- {e}_1) & = \lambda_1 {e}_1(1- {e}_2)\nonumber \\&
                      = \lambda_2  {e}_2(1- {e}_3)   \nonumber \\ & \vdots \nonumber \\
                    &= \lambda_{n-1} {e}_{n-1} (1- {e}_n) \nonumber \\& =\lambda_n  {e}_n.
\end{align}
Denoting the \emph{steady-state translation rate} by
\be \label{eq:defr}
R:=\lambda_n  {e}_n
\ee
yields
\begin{align}\label{eq:rall}
R=\lambda_i e_i(1-e_{i+1}), \quad i=0,\dots,n,
\end{align}
where $e_0:=1$ and $e_{n+1}:=0$. Thus
\be\label{eq:ei+1}
e_{i+1}=1-\frac{R}{\lambda_i e_i}=\frac{R_i-R}{R_i},
\ee
where $R_i:=\lambda_i e_i$.
Since $e_{i+1}\in(0,1)$,   $0< R < R_i$ for all $i=1,2,\dots,n-1$. This means that the steady-state occupancy level~$e_{i}$
is the normalized difference between~$R_i$ and the steady-state translation rate.

Using~\eqref{eq:defr}, Eq.~\eqref{eq:rfm} becomes
\begin{align}\label{eq:list}
                             {e}_n & = R/\lambda_n, \nonumber  \\
                             {e}_{n-1} & = R / (\lambda_{n-1} (1- {e}_n) ),  \nonumber\\
                            & \vdots \nonumber \\
                             {e}_{2} & = R / (\lambda_{2} (1- {e}_3) ), \nonumber\\
                             {e}_{1} & = R / (\lambda_{1} (1- {e}_2) ),
\end{align}
and \be \label{eq:also}
                             {e}_1= 1-R/ \lambda_0 .
\ee

Combining~\eqref{eq:list} and~\eqref{eq:also} provides an elegant
\emph{finite continued fraction}~\cite{waad} expression for~$R$:

\begin{align} \label{eq:cf}
                0&= 1-\cfrac{R/ \lambda_0 }
                                  {  1-\cfrac{R / \lambda_1}
                                  {1-\cfrac{R / \lambda_2}{\hphantom{aaaaaaa} \ddots
                             \genfrac{}{}{0pt}{0}{}
                             {1-\cfrac{R/\lambda_{n-1}}{1-R/ \lambda_n.}} }}}
\end{align}
Note that this equation admits several solutions for~$R$, however, we are interested only in the unique feasible solution, i.e. the solution corresponding to~$e \in \Int(C^n)$.
Note also that~\eqref{eq:cf} implies that
 \be\label{eq:rishomog}
R(c \lambda_0,\dots,c \lambda_n)= c R (  \lambda_0,\dots,  \lambda_n)  ,\quad \text{for all }c>0,
 \ee
  that is,~$R$ is a \emph{homogenous  function} of degree one.

 It is well-known that continued fractions are related to tridiagonal matrices~\cite{wall_contin_frac,sturm_tridiagnal}.
 Using this, Ref.~\cite{rfm_max}  provided a linear-algebraic representation of the mapping from the rates~$\lambda:=\begin{bmatrix}
 \lambda_0,\dots,\lambda_n\end{bmatrix}'$ to the steady state translation rate~$R$.
\begin{Theorem}\label{thm:linear_rep} \cite{rfm_max}
Given~$\lambda \in \Int(\R^{n+1}_+)$,
define a~$(n+2)\times(n+2)$ symmetric irreducible Jacobi matrix $A=A(\lambda)$ by
\be\label{eq:matrix_A}
\small
                A:= \begin{bmatrix}
 0 &  \lambda_0^{-1/2}   & 0 &0 & \dots &0&0 \\
\lambda_0^{-1/2} & 0  & \lambda_1^{-1/2}   & 0  & \dots &0&0 \\
 0& \lambda_1^{-1/2} & 0 &  \lambda_2^{-1/2}    & \dots &0&0 \\
 & &&\vdots \\
 0& 0 & 0 & \dots &\lambda_{n-1}^{-1/2}  & 0& \lambda_{n }^{-1/2}     \\
 0& 0 & 0 & \dots &0 & \lambda_{n }^{-1/2}  & 0
 \end{bmatrix}.
\ee
Then the eigenvalues of~$A$ are real and distinct, and if we order them as
$
\zeta_1<\dots<\zeta_{n+2}
$
then~$\zeta_{n+2}=R^{-1/2}$.
\end{Theorem}
Note that~$A$ is a  (componentwise) nonnegative matrix, so~$\zeta_{n+2}$ is also
the Perron root of~$A$, denoted~$\rho(A)$.

Recently, the RFM was analyzed using tools from systems and control theory.
Ref.~\cite{RFM_feedback} has considered the~RFM as a control system with input~$u(t)=\lambda_0(t)$
and output~$y(t)=R(t)$. This turns out to be a monotone control system,
as defined in~\cite{mcs_angeli_2003}.
 \emph{Ribosome recycling}
(see, e.g.,~\cite{recycle2013} and the references therein),
has been modeled by closing the loop with a positive linear feedback.
  It has been shown that the closed-loop system admits a unique globally asymptotically stable equilibrium point~\cite{RFM_feedback}.
 In~\cite{RFM_entrain}, contraction theory (see, e.g., \cite{LOHMILLER1998683,three_gen_cont,sontag_contraction_tutorial})
 has been used to show that the state-variables and the translation rate in the RFM \emph{entrain} to periodically time-varying initiation and/or elongation rates. This provides a computational framework for studying entrainment to a periodic excitation, like
 those involved in the cell-division process,  at the translation level.

 \subsection{Maximizing  the Translation Rate}
An important problem in both systems biology and biotechnology is to
maximize  the protein production rate, given the limited
biomolecular budget.
Ref.~\cite{rfm_max} formulated this in the context of the RFM
as  the following  optimization problem.
 \begin{Problem}\label{prob:max}
Fix the parameters b,$w_0,w_1,\dots,w_n  >0$.  Maximize~$R =R(\LMD)$, with respect to its parameters $\LMD$, subject to the constraints:
\begin{align}\label{eq:constraint}
\sum_{i=0}^n w_i\lambda_i  & \leq  b, \\
\LMD &\geq 0.\nonumber
\end{align}
\end{Problem}
In other words, maximize the translation rate   given an affine constraint that
 takes into account all the rates $\lambda_i$, $i=0,\dots,n$. This is related to factors such as the abundance of intracellular ribosomes, initiation factors, intracellular tRNA molecules, and elongation factors.  For example, all tRNA molecules are transcripted by the same transcription factors~(TFIIIB) and by RNA polymerase III.
The values $w_i$, $i=0,\dots,n$, can be used to provide different weighting to the different rates.  It has been shown in~\cite{rfm_max} that the optimal solution
always satisfies~$\sum_{i=0}^n w_i\lambda_i^*= b$.
Of course, by scaling the~$w_i$s we may always assume that~$b=1$.

By Theorem~\ref{thm:linear_rep},   Problem~\ref{prob:max} is equivalent to the following
eigenvalue minimization  problem.
 \begin{Problem}\label{prob:min_perron}
Let $b,w_0,w_1,\dots,w_n  >0$ be as in Problem~\ref{prob:max}.
 Consider the matrix $A$ in~\eqref{eq:matrix_A}. Minimize~$\rho(A)$ with respect to the parameters $\LMD$, subject to the constraints in~\eqref{eq:constraint}.
\end{Problem}

The  two problems above are equivalent in the sense that~$\lambda^*$ is a solution of one problem if and only if
it is a  solution of the second. Also, for~$A^*:=A(\lambda^*)$ we have
$\rho(A^*)=(R^*)^{-1/2}$.

 The next result from~\cite{rfm_max} shows that Problem~\ref{prob:min_perron} (and thus Problem~\ref{prob:max}) admits several desirable properties.
 Let~$\Int(\R^k_+)$ denote the interior of~$\R^k_+$ i.e. the set~$\{x\in \R^k: x_i>0,\; i=1,\dots, k \}$.
 \begin{Theorem}~\cite{rfm_max}
 The solution~$\lambda^*=\begin{bmatrix} \lambda_0^*,\dots,\lambda_n^*\end{bmatrix}' $ of Problem~\ref{prob:max}
 satisfies~$\lambda_i^*>0$ for all~$i$. Furthermore, the function~$R=R(\lambda)$ is strictly concave on~$\Int(\R^{n+1}_+)$,
 so Problem~\ref{prob:max} is a convex optimization problem, and the solution~$\lambda^*$ is unique and can be found using numerical algorithms that scale well with~$n$.
 The optimal steady-state production rate~$R^*:= R(\LMDSTAR)$ satisfies
\be\label{eq:opt_R}
R^*=\frac{(\lambda_0^*)^2}{\lambda_0^*+\frac{w_1}{w_0}\lambda_1^*}.
\ee
 \end{Theorem}

\subsubsection{Maximization subject to a homogeneous constraint}
It is   interesting to consider the
case where all the weights~$w_i$ in Problem~\ref{prob:max} are equal.
We refer to this  as the \emph{homogeneous constraint} case.
 Indeed, in this case the weights
give equal preference to all the rates, so
if the corresponding
optimal solution  satisfies~$\lambda_i^* > \lambda_j^*$ for some~$i,j$ then this implies
that, in the context of maximizing~$R$,~$\lambda_i$ is ``more important'' than~$\lambda_j$.
By~\eqref{eq:rishomog}, we may assume in this case, without loss of generality, that
\be \label{eq:homog_comnst}
w_0=\dots=w_n=b=1.
\ee

\begin{Example}\label{exp:n7_hom}
Consider Problem~\ref{prob:max} for an RFM with~$n=7$
 and  the homogenous constraint~\eqref{eq:homog_comnst}.
 Fig.~\ref{fig:rfm_n7_l} depicts the optimal values $\lambda_i^*$, $i=0,\dots,7$, computed using a simple search algorithm (that is guaranteed to converge for convex optimization problems). It may be seen
  that the~$\lambda_i^*$s are symmetric, i.e.
  $\lambda_i^*=\lambda_{7-i}^*$, and that
   they increase towards the center of the chain.
   Fig.~\ref{fig:rfm_n7_e} depicts the corresponding optimal values $e_i^*$.
   It may be seen that  the steady-state occupancy levels
    strictly decreases along the chain.
\begin{figure}[t]
  \begin{center}
  \includegraphics[width= 7cm,height=6cm]{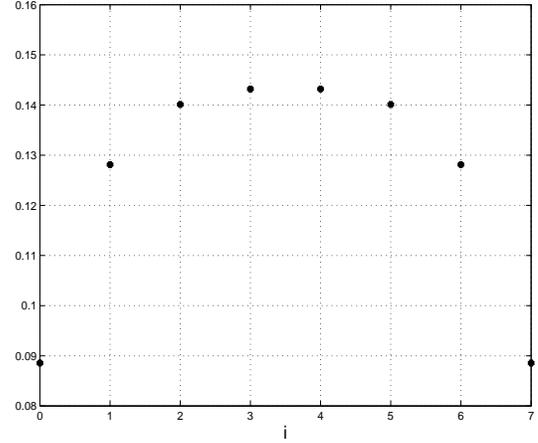}
  \caption{  $\lambda_i^*$ as a function of~$i$
  for an RFM with $n=7$ and the homogenous constraint~\eqref{eq:homog_comnst}.}\label{fig:rfm_n7_l}
  \end{center}
\end{figure}
\begin{figure}[t]
  \begin{center}
  \includegraphics[width= 7cm,height=6cm]{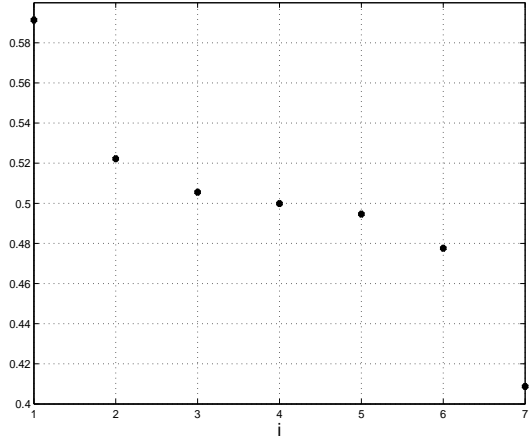}
  \caption{ $e_i^*$ as a function of~$i$
  for an RFM with $n=7$ and the homogenous constraint~\eqref{eq:homog_comnst}.}\label{fig:rfm_n7_e}
  \end{center}
\end{figure}
\end{Example}

In this paper, we analyze  the properties
 of the optimal rates~$\lambda^*$, the corresponding
 steady-state occupancies~$e^*$, and translation rate~$R^*$.
  The next section describes our main results.
 Section~\ref{sec:conc} describes some of the
  biological implications of the theoretical results,
 and describes possible directions for further research.
  To streamline the presentation, all
 the proofs are placed in the Appendix.

\section{Main results}
We  begin by stating several results that hold in general for Problem~\ref{prob:max}.
These results are of independent interest, and will also be used to analyze the
specific case of the homogeneous constraint below.

\subsection{Sensitivity at the Optimal Rates}
Given~$\lambda^*\in \Int(\R^{n+1})$, pick~$i,j \in \{0,\dots,n\}$, with~$i\not =j$,
and consider the vector~$\tilde \lambda$ defined by
\[
            \tilde \lambda_k:=\begin{cases}
                                  \lambda_i^*+\frac{\varepsilon}{w_i}   ,  & k=i,\\
                                  \lambda_j^*-\frac{\varepsilon}{w_j}   ,  & k=j,\\
                                  \lambda_k^*   ,  & \text{otherwise} ,\end{cases}
\]
where~$|\varepsilon|$ is sufficiently small  so that
$\tilde \lambda_k>0$ for all~$k$. Note that~$\sum_{k=0}^n w_k \tilde \lambda_k=
\sum_{k=0}^n w_k   \lambda_k^*=b$. Then
\begin{align*}
                    \tilde R & :=R(\tilde \lambda) \\
                             &=   R(\lambda^*)+\frac{\varepsilon}{w_i}     \frac{\partial  R(\lambda^*)}{\partial \lambda_i} - \frac{\varepsilon}{w_j}     \frac{\partial  R(\lambda^*)}{\partial \lambda_j}
                              +o(\varepsilon).
\end{align*}
Using the fact that~$R^*>\tilde R$ for all~$\varepsilon \in \R\setminus\{0\}$ (with~$|\varepsilon|$  sufficiently small)
implies that
\be\label{eq:eq_sese}
\frac{1}{w_i} \frac{\partial  R(\lambda^*)}{\partial \lambda_i}=\frac{1}{w_j} \frac{\partial  R(\lambda^*)}{\partial \lambda_j}  ,\quad \text{for all } i,j.
\ee
In other words, the weighted sensitivities at the optimal values are all   equal.
  This result has already been derived in~\cite{rfm_max}.
Here we give a slightly stronger result that provides a closed-form  expression for the weighted sensitivities.
\begin{Proposition}\label{prop:opt_perron_sens}
Consider Problem~\ref{prob:max}. The weighted sensitivities  at the optimal parameter values satisfy
\be\label{eq:opt_r_deriv}
\frac{1}{w_i} \frac{\partial}{\partial \lambda_i} R (\lambda^*) = \frac{  R (\lambda^*)}{b},\quad i=0,\dots, n.
\ee
\end{Proposition}

\begin{Remark}\label{rem:sensi}
For the case of a homogeneous constraint, this yields~$  \frac{\partial}{\partial \lambda_i} R (\lambda^*) =  R (\lambda^*) /b$, $i=0,\dots, n$,
so in particular all the sensitivities at the optimal solution are equal.
This is reasonable, as otherwise it would be possible to find a better solution
by placing a larger [smaller] rate~$\tilde \lambda_i$ [$\tilde \lambda_j]$
at site~$i$ $[j]$ that has a higher [lower] sensitivity, while preserving the total bound on the rates.
This also implies that to first-order in~$\varepsilon$,
\begin{align*}
            R(\lambda^*+\varepsilon e^k)&=R(\lambda^*)+\varepsilon \frac{\partial}{\partial \lambda_k} R (\lambda^*)\\
&=            R(\lambda^*)+\varepsilon   R (\lambda^*),
\end{align*}
where~$e^k$ is the vector with~$1$ in entry~$k$, and $0$ elsewhere.
\end{Remark}

Note that Prop.~\ref{prop:opt_perron_sens} implies that the optimal
solution of Problem~\ref{prob:min_perron}
satisfies
\be\label{eq:opt_p_deriv}
 \frac{1}{w_i}\frac{\partial}{\partial \lambda_i} \rho(A^*)  =
 -\frac{1}{2b}\rho(A^*), \quad i=0,\dots, n.
\ee

 \begin{Example}
Consider the very simple case
of an RFM with $n=1$.
  Solving~\eqref{eq:cf} yields $R=\frac{\lambda_0 \lambda_1}{\lambda_0 + \lambda_1}$. Using the constraint $w_0\lambda_0+w_1\lambda_1=b$
  yields~$R=\frac{\lambda_0(b-w_0 \lambda_0 )}{  (w_1-w_0)\lambda_0+b } $.
  The unique    maximizing value is~$\lambda_0^*=\frac{b}{w_0+\sqrt{w_0 w_1}}$,
  so~$\lambda_1^*= \frac{b}{w_1+\sqrt{w_0 w_1}}$, and
  \begin{align*}
        R^*&=R(\lambda_0^*,\lambda_1^*)\\&=\frac{b}{(\sqrt{w_0}+\sqrt{w_1})^2}.
  \end{align*}
On the other-hand, the derivatives are
\[
            \frac{\partial R}{\partial \lambda_0}=\frac{\lambda_1^2}{(\lambda_0+\lambda_1)^2}  ,
 \quad
 \frac{\partial R}{\partial \lambda_1}=\frac{\lambda_0^2}{(\lambda_0+\lambda_1)^2}     ,
\]
and substituting the optimal values yields~$\frac{\partial R(\lambda^*)}{\partial \lambda_i}=w_i R^*/b$, $i=0,1$.
\end{Example}

\subsection{Optimal Steady-State Occupancies}

Let~$e_i^*=e_i(\LMDSTAR)$ denote the steady-state occupancy levels corresponding to the optimal rates.
Our first result relates these  occupancy levels  to the  optimal rates.

\begin{Proposition}\label{prop:lambda_ratio}
The  steady-state occupancy levels corresponding to the optimal rates satisfy
\be\label{eq:lambda_ratio}
\frac{\lambda_{i+1}^*}{\lambda_i^*}=\frac{w_i}{w_{i+1}}\frac{e_{i+1}^*}{1-e_{i+1}^*},\quad i=0,\dots,n-1.
\ee
\end{Proposition}

Prop.~\ref{prop:lambda_ratio} implies that given the optimal rates~$\lambda_i^*$
 the corresponding steady-state occupancies can be determined   via
 \be\label{eq:e_lambda_ratio}
e_{i+1}^* = \left(1+\frac{\lambda_i^*}{\lambda_{i+1}^*}\frac{w_i}{w_{i+1}}   \right)^{-1},
\ee
 instead of by solving~\eqref{eq:ep}.

The~$e^*_i$s thus satisfy two sets of equations: the set given in Prop.~\ref{prop:lambda_ratio},
and the set based on the RFM steady-state equation~\eqref{eq:ep}, that is,
\[
  e_i^*  =\frac{  \lambda_{i+1}^*}{\lambda_i^*}  \frac{ e_{i+1}^*}{1-e_{i+1}^*}  (1-e_{i+2}^*),\quad i=0,\dots,n-1.
\]

Combining these sets of equations yields the following.
\begin{Corollary}
The optimal occupancies satisfy
\be\label{eq:combined}
  e_i^*  =\frac{  w_{i }}{w_{i+1}} \left( \frac{ e_{i+1}^*}{1-e_{i+1}^*}\right)^2  (1-e_{i+2}^*),\quad i=0,\dots,n-1,
\ee
with~$e_0^*:=1$ and~$e^*_{n+1}:=0$.
\end{Corollary}

\begin{Example}
Consider Problem~\ref{prob:max} for an  RFM with $n=3$ and the homogeneous constraint~\eqref{eq:homog_comnst}. In this case,~\eqref{eq:combined} yields
\begin{align*}
1&=\left( \frac{ e_1^*}{1-e_1^*} \right )^2 (1-e_2^*), \\
e_1^*&=\left( \frac{ e_2^*}{1-e_2^*} \right )^2 (1-e_3^*), \\
e_2^*&= \left(  \frac{ e_3^*} {1-e_3^*}\right )^2.
\end{align*}
Solving this yields
\be\label{eq:esn=3}
e^*=\begin{bmatrix} 2-\sqrt{2},1/2,\sqrt{2}-1 \end{bmatrix}'.
\ee
Now applying~\eqref{eq:lambda_ratio}
 and using the fact that~$\lambda_0^*+\dots+ \lambda_3^*=1$,
yields
\be\label{eq:lamstarn=3}
\lambda^*=\begin{bmatrix}  \frac{2\sqrt{2}-1}{\sqrt{7}},
\frac{4-\sqrt{2} }{ {7}},
\frac{4-\sqrt{2} }{ {7}},
\frac{2\sqrt{2}-1}{\sqrt{7}}
 \end{bmatrix}'.
 \ee
\end{Example}

\begin{Remark}
An important question is how many~$\lambda_i^*$ values are needed in order to
uniquely determine~$R^*$? It  follows from~\eqref{eq:opt_R}
 that knowing~$\lambda_0^*$ and~$\lambda_1^*$ is enough.
Also, it follows from~\eqref{eq:e_lambda_ratio} with~$i=n-1$ that
\begin{align*}
                R^*&=\lambda_n^* e_n^* \\
                   &=\frac{(\lambda_n^* )^2}{\lambda_n^* +\frac{ w_{n-1}}{w_n} \lambda_{n-1}^* }.
\end{align*}
Thus, knowing either the first or the last two optimal rates
is enough to uniquely determine~$R^*$.
\end{Remark}

\subsection{  Homogenous Constraint  }
As noted above, it is interesting to
consider Problem~\ref{prob:max} with
 the homogeneous constraint~\eqref{eq:homog_comnst}.
The following result proves that in this case the optimal solution
$\lambda^*$ and the corresponding~$e^*$ always
have the structure depicted in Figs.~\ref{fig:rfm_n7_l} and~\ref{fig:rfm_n7_e}.
\begin{Proposition}\label{prop:lambda_ratio_hom}
Consider Problem~\ref{prob:max} with the
 homogenous constraint~\eqref{eq:homog_comnst}. Then
the following properties hold.
\begin{itemize}
\item The optimal rates satisfy~$\lambda_0^*<\lambda_1^*<\dots<\lambda_{\lfloor  \frac{n}{2} \rfloor}^*$, and
\be\label{eq:symlamopt}
 \lambda_i^*=\lambda_{n-i}^* ,\quad  i=0,\dots,n.
 \ee
\item The corresponding steady-state occupancies satisfy
\be\label{eq:eismono}
e_i^*=1-e_{n-i+1}^*,\quad i=1,\dots,n.
\ee
If $n$ is even then
\be\label{eq:niseven}
          e_1^*> \dots >e_{\frac{n}{2}}^*>\frac{1}{2}>  e_{\frac{n}{2}+1}^*> \dots>  e_{n}^*,
\ee
and if $n$ is odd then
\be\label{eq:eisnodd}
          e_1^*> \dots > e_{\frac{n-1}{2}}^*  > e_{\frac{n+1}{2}}^*=\frac{1}{2}>    e_{\frac{n+2}{2}}^*> \dots>  e_{n}^*.
\ee
\end{itemize}
\end{Proposition}

Note that~\eqref{eq:eismono}
implies that $\sum_{i=1}^n e_i^*=n/2$.
Note also  that   the  results in~\eqref{eq:esn=3}  and~\eqref{eq:lamstarn=3} agree of
course with the results in Prop.~\ref{prop:lambda_ratio_hom}.


\section{Discussion}\label{sec:conc}
Gene translation is known to be one of the most energy
consuming processes in the cell. Thus, it is reasonable to assume that this process evolved such that the protein production rate of highly-expressed genes is optimized given the limited cell resources.
Maximizing the translation rate is also important in gene cloning for biotechnological applications.

The RFM is a deterministic mathematical model for translation-elongation obtained
via a mean-field approximation of a fundamental model from non-equilibrium statistical  mechanics called TASEP. The RFM includes~$n+1$
  positive parameters: the   initiation rate~$\lambda_0$, and the elongation rates~$\lambda_1,\dots,\lambda_n$.

It is possible to formulate
the  problem of optimizing the steady-state
translation rate~$R$, subject to the limited biomolecular budget,
as a constrained  optimization problem using the RFM. This problem has several desirable properties that follow
from the fact that~$R$ is a strictly concave function of the rates $\lambda_i$, $i=0,\dots,n$~\cite{rfm_max}.

In this paper, we   analyzed   the optimal vector of initiation and elongation
rates~$\lambda^*$, and the corresponding
steady-state occupancies (ribosomal  densities)~$e^*$.
Our results show that  for a constraint that gives equal weighting to all the rates,
$\lambda^*$ has a special structure:
the rates~$\lambda^*_i$  are symmetric with respect to the center of the chain, and strictly decrease
as we move towards the ends of the chain. This holds for every dimension~$n$.
The reason for this structure is the particle-hole symmetry, and the fact
that sites at the center of the chain have a large number of neighboring  sites.
These results agree with the so called ``edge-effect" in TASEP (see, for example,~\cite{toward_prod_rates} and~\cite{PhysRevE.76.051113}), i.e. the fact that the output rate is less sensitive to
 the rates close to the edges of the chain.

Since the   optimal rates close
to the two ends of the chain are relatively small,
these rates may be considered as the
 limiting factors of the translation process.
 Yet, this intuitive interpretation is wrong.
  Indeed, at the optimal solution all the sensitivities to a change in any of the rates are equal
  (see Remark~\ref{rem:sensi}), so all the rates limit the translation rate to the \emph{same extent}.
  The rates towards the center of the chain have a higher effect
 on~$R$ and, therefore, the optimal solution includes lower rates at the ends of the chain.
 An important open problem in
 gene translation is what is the dominant gene translation regime:
 some studies claim that initiation is the rate limiting step~\cite{Kudla2009},
  while others claim that the elongation  step  is also
     rate limiting~\cite{Supek2010,Tuller2010}.
Our results suggest that in addressing this question one must take into account not only
the values of the initiation and elongation rates, but also
the \emph{sensitivity} of the production rate with respect to these rates.

An interesting topic for further research  is
maximizing  the protein production rate when some of the rates are fixed.
For example, suppose that we fix one or more of the rates to small values so that
they form a bottleneck in the chain. Then optimizing over the other rates provides information on
how to overcome the decrease due to the  bottlenecks in an optimal manner.
For example, consider the RFM with~$n=4$ and the homogeneous constraint~\eqref{eq:homog_comnst}.
The optimal rates are
\be\label{eq:add_opt_4}
\lambda^*=\begin{bmatrix} 0.1559, 0.2243, 0.2396, 0.2243, 0.1559\end{bmatrix}' ,
\ee
 and $R^*=0.0639$.
Now suppose that we fix~$\lambda_2=0.02$, and optimize over the other rates
subject to  the constraint~$\lambda_0+\lambda_1+ \lambda_3+\lambda_4\leq 1-0.02=0.98$.
Now the optimal solution is
\[
\tilde \lambda=\begin{bmatrix}0.0935, 0.3965, 0.02, 0.3965, 0.0935 \end{bmatrix}' ,
\]
  and~$\tilde R=0.0178$.
  Comparing this to~\eqref{eq:add_opt_4}, we see
  that~$\tilde \lambda_i>\lambda^*_i$, $i=1,3$. Indeed, these rates must increase in order to compensate
   for the forced slow rate at site~$2$. Note that although~$\tilde \lambda_2 /\lambda_2^*=0.083$,
   $\tilde R/R^*=0.2786$.
   Thus, the optimal solution is able to compensate, to some extent,
   for the  drastic reduction in the elongation rate.

\begin{figure}[t]
  \begin{center}
  \includegraphics[width= 7cm,height=6cm]{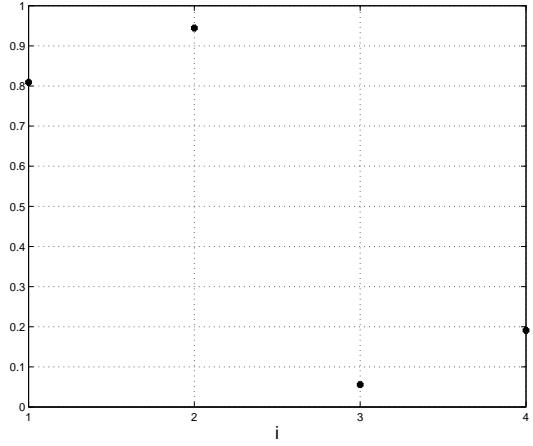}
  \caption{  $\tilde e_i $ as a function of~$i$
  for an RFM with $n=4$.}\label{fig:n4_forced}
  \end{center}
\end{figure}

  Fig.~\ref{fig:n4_forced} depicts the occupancy levels $\tilde e_i$, $i=1,\dots,4$, corresponding to~$\tilde \lambda$.
  Now of course the properties in Prop.~\ref{prop:lambda_ratio_hom} no longer hold.
  As expected, forcing~$\lambda_2$ to a low value yields a high [low] value for~$\tilde e_2$ [$\tilde e_3$].
\section*{Appendix: Proofs}
{\sl Proof of Prop.~\ref{prop:opt_perron_sens}.}
By Euler's theorem for homogeneous functions,
\begin{align*}
            R(\lambda)&=\sum_{i=0}^n \lambda_i \frac{\partial R(\lambda)}{\partial \lambda_i} \\
                      &=\sum_{i=0}^n \lambda_i w_i \frac{1}{w_i} \frac{\partial R(\lambda)}{\partial \lambda_i} .
\end{align*}
 Substituting~$\lambda=\lambda^*$ and using~\eqref{eq:eq_sese} yields
\begin{align*}
R(\lambda^*)&=\sum_{i=0}^n \lambda_i^* w_i \frac{1}{w_i} \frac{\partial R(\lambda^*)}{\partial \lambda_i}\\
 &=\frac{1}{w_1} \frac{\partial R(\lambda^*)}{\partial \lambda_1} \sum_{i=0}^n \lambda_i^* w_i  \\
 &= \frac{1}{w_1} \frac{\partial R(\lambda^*)}{\partial \lambda_1}  b,
\end{align*}
and this  completes the proof of Prop.~\ref{prop:opt_perron_sens}. \IEEEQED


{\sl Proof of Prop.~\ref{prop:lambda_ratio}.}
 For~$\lambda\in\Int(\R^{n+1}_+)$
  and~$i \in \{0,\dots,n\}$, define
\begin{align}\label{eq:G}
G_{n-i}(z ):=1-\cfrac{z/ \lambda_i }
                                  {  1-\cfrac{z / \lambda_{i+1}}
                                  {1-\cfrac{z / \lambda_{i+2}}{\hphantom{aaaaaaa} \ddots
                             \genfrac{}{}{0pt}{0}{}
                             {1-\cfrac{ z/\lambda_{n-1}}{1-z/ \lambda_n,}} }}}.
\end{align}
 Note that~\eqref{eq:cf} implies that
 \be\label{eq:gatval}
 G_n(R(\lambda )  )=0,\quad \text{for all } \lambda\in\Int(\R^{n+1}_+),
 \ee
 and that~\eqref{eq:list} and~\eqref{eq:also} yield
 \be\label{eq:gn1e}
 G_{n-i}(R(\lambda ))=1-e_i(\lambda),
 \ee
for all $ \lambda\in\Int(\R^{n+1}_+),\; i=0,\dots,n$.

We require the following result.
\begin{Proposition}\label{prop:dergb}
For all~$i=0,\dots,n$,
\begin{align*}
\frac{\partial G_n}{\partial\lambda_i} (z)
&=\frac{z^{i+1}}{   \left(\prod_{\ell=0}^{i-1}\lambda_\ell \right )  \left(\prod_{\ell=1}^{i }G^2_{n-\ell}(z)\right) \lambda_i^{2}G_{n-(i+1) } (z)} .
\end{align*}
\end{Proposition}

{\sl Proof of Prop.~\ref{prop:dergb}.}
Pick~$k\in\{0,1\dots,n-1\}$.
By~\eqref{eq:G},
 $G_{n-k}(z)=1-\frac{z/\lambda_k}{G_{n-(k+1)}(z)}$  for all $k$.
Therefore,
\be\label{eq:part_gn_1}
\frac{\partial G_{n-k}}{\partial\lambda_j}=
\frac{z}{\lambda_k G_{n-(k+1)}^2} \frac{\partial G_{n-(k+1)}}{\partial\lambda_j}, \quad \text{for all } j\not =k,
\ee
and
\be\label{eq:part_gn_2}
\frac{\partial G_{n-k}}{\partial\lambda_k}=\frac{z }{\lambda_k^{2} G_{n-(k+1)}}.
\ee
Thus,
\begin{align*}
\frac{\partial G_n}{\partial\lambda_i} &= \frac{z}{\lambda_0 G_{n-1}^2} \frac{\partial G_{n-1}}{\partial\lambda_i}\\
&= \frac{z}{\lambda_0 G_{n-1}^2}  \frac{z}{\lambda_1 G_{n-2}^2} \frac{\partial G_{n-2}}{\partial\lambda_i} \\
&\vdots\\
&=\frac{z^i}{   (\prod_{\ell=0}^{i-1}\lambda_\ell )( \prod_{\ell=1}^{i }G^2_{n-\ell})}   \frac{\partial G_{n-i}}{\partial\lambda_i} , 
\end{align*}
and using~\eqref{eq:part_gn_2} completes the proof of Prop.~\ref{prop:dergb}.~\IEEEQED

We can now proceed with the proof of Prop.~\ref{prop:lambda_ratio}.
  It follows from~\eqref{eq:gatval} that
\begin{align*}
0&= \frac{d G_n}{d\lambda_i}(R)\\
& =\frac{\partial G_n(R)}{\partial R}\frac{\partial R}{\partial \lambda_i}+\frac{\partial G_n(R)}{\partial \lambda_i},
\end{align*}
 for all~$i \in \{0,\dots,n\}, \lambda\in\Int(\R^{n+1}_+)$.
Substituting the optimal parameter values and using~\eqref{eq:opt_r_deriv} yields
\be\label{eq:partial_G_ratio}
 0=\frac{\partial G_n(R^* )}{\partial R}
\frac{ w_i R^* }{b} +\frac{\partial G_n(R^* )}{\partial \lambda_i},\quad
i=0,\dots,n.
\ee
Combining this with Prop.~\ref{prop:dergb} and~\eqref{eq:gn1e} yields
\[
 \frac{\partial G_n(R^* )}{\partial R}
= \frac{-b(R^*)^{i }}{  w_i  \left(\prod_{\ell=0}^{i }\lambda^*_\ell \right )  \left(\prod_{\ell=1}^{i } (1-e^*_\ell)^ 2\right) \lambda_i^*  (1-e^*_{i+1})  },
\]
  for all~$i=0,\dots,n$. Pick~$j \in \{0,\dots,n-1\}$. Then
\begin{align*}
 &\frac{ (R^*)^{j }}{  w_j  \left(\prod_{\ell=0}^{j }\lambda^*_\ell \right )  \left(\prod_{\ell=1}^{j } (1-e^*_\ell)^ 2\right) \lambda_j^*  (1-e^*_{j+1})  }
 \\&=\frac{ (R^*)^{j+1 }}{  w_{j+1}  \left(\prod_{\ell=0}^{j+1 }\lambda^*_\ell \right )  \left(\prod_{\ell=1}^{j+1 } (1-e^*_\ell)^ 2\right) \lambda_{j+1}^*  (1-e^*_{j+2})  } .
\end{align*}
Simplifying this and using the fact that~$R^*=\lambda_{j+1}^* e_{j+1} ^*(1-e_{j+2}^*)$
(see~\eqref{eq:rall}) yields~\eqref{eq:lambda_ratio}.~\IEEEQED


{\sl Proof of Prop.~\ref{prop:lambda_ratio_hom}.}
Let~$x$ denote the coordinates of an~RFM with dimension~$n$
and rates~$\lambda=\begin{bmatrix}\zeta_0,\zeta_1,\dots,\zeta_n \end{bmatrix}'$. Define~$\tilde x_i(t):=1-x_{n -i+1}(t)$, $i=1,\ldots,n$.
It is straightforward to verify that the dynamics
of the~$\tilde x$-system is just that of the RFM, but with rates~$\tilde \lambda=\begin{bmatrix} \zeta_n,\zeta_{n-1},\dots,\zeta_0\end{bmatrix}'$.
Since~$x_i$ converges to~$e_i$, $\tilde x_i$ converges to~$1-e_{n-i+1}$.
This proves the following.
\begin{Fact}\label{fact:particle-hole}
Consider two RFMs  with dimension n, one  with rates $\lambda=\begin{bmatrix}\zeta_0,\zeta_1,\dots,\zeta_n \end{bmatrix}'$ and steady-state $e$, and the second with rates $\tilde \lambda=\begin{bmatrix} \zeta_n,\zeta_{n-1},\dots,\zeta_0\end{bmatrix}'$ and steady-state $\tilde e$. Then
\be\label{eq:phs2}
\tilde e_i=1- e_{n-i+1}, \quad   i=1,\dots,n.
\ee
\end{Fact}

Intuitively  speaking,
  the unidirectional flow of particles   along the chain
   may also be interpreted as the flow   of ``holes''  in the opposite direction,
 and this yields Fact~\ref{fact:particle-hole}.
 In the TASEP literature, this is known as
  the \emph{particle-hole symmetry} (see, e.g.,~\cite{PhysRevE.71.011103}).

Note that~\eqref{eq:phs2}
 implies in particular that
\begin{align*}
                        R&=\zeta_n e_n \\
                         &=\tilde \lambda_0 (1-\tilde e_{ 1})\\
                         &=\tilde R,
\end{align*}
where the last step follows from~\eqref{eq:rall}. Thus,
\be\label{eq:phs1}
R(\zeta_0,\zeta_1,\dots,\zeta_n)=R(\zeta_n,\zeta_{n-1},\dots,\zeta_0).
\ee

 Let~$\lambda^*$ denote the optimal solution for Problem~\ref{prob:max} with the homogeneous constraint.
Since $w_i=w_{n-i}$ for all $i=0,\dots,\lfloor n/2 \rfloor$,
the vector~$\tilde \lambda^*:=\begin{bmatrix} \lambda^*_n ,\lambda^*_{n-1} , \dots, \lambda^*_0\end{bmatrix}'$
  satisfies the constraint~$\sum_{i=0}^n w_i \tilde \lambda_i^*=b$. By~\eqref{eq:phs1}, $\tilde \lambda^*$ is also
an optimal solution. By uniqueness  of the optimal solution,~$\tilde \lambda^*=\lambda^*$.
This proves~\eqref{eq:symlamopt}, and using Fact~\ref{fact:particle-hole} proves~\eqref{eq:eismono}.

To prove that the~$\lambda_i^*$s increase as we move towards the center of the chain,
  note that by~\eqref{eq:lambda_ratio},
    \be\label{eq:adedat}
\frac{\lambda_1^*}{\lambda_0^*}=\frac{e_1^*}{1-e_1^*}.
\ee
On the other-hand~\eqref{eq:ep} gives~$\frac{\lambda_1^*}{\lambda_0^*}=\frac{1-e_1^*}{e_1^*}\frac{1}{1-e_2^*}$, so
\[
\left(\frac{\lambda_1^*}{\lambda_0^*}\right)^2=\frac{1}{1-e_2^*},
\]
and since $e_2^*\in(0,1)$,
\be\label{eq:l0l1_ratio}
\lambda_0^*<\lambda_1^*,
\ee
and~\eqref{eq:adedat} implies that
\be\label{eq:e112}
e_1^*>1/2.
\ee
Consider first the case where~$n$ is even. Let~$m:=n/2$.
We need to show that
\be\label{eq:ntsq}
     \lambda^*_0<   \lambda_1^*  <\dots  < \lambda_m^*.
\ee
Seeking a contradiction, assume that
\be\label{eq:l2l1}
\lambda_{m-1}^* \geq \lambda_m^*.
\ee
Then~\eqref{eq:lambda_ratio} gives~$e_m^*\leq 1/2$. Now~\eqref{eq:ep} and~\eqref{eq:eismono}
 yield
\begin{align*}
                        e^*_{m-1}&=\frac{\lambda^*_m}{\lambda^*_{m-1}}
                        \frac{e^*_m}{1-e^*_m}(1-e^*_{m+1})\\
                        &\leq 1\times 1\times e^*_{n-m}\\
                        &=e^*_{m},
\end{align*}
so~$e^*_{m-1}\leq 1/2$. Combining this with~\eqref{eq:lambda_ratio} yields
$ \lambda_{m-2}^* \geq \lambda_{m-1}^* $.
Now~\eqref{eq:ep} and~\eqref{eq:eismono}
 yield
\begin{align*}
                        e^*_{m-2}&=\frac{\lambda^*_{m}}{\lambda^*_{m-2}}
                        \frac{e^*_{m }}{1-e^*_{m-1}}(1-e^*_{m+1 })\\
                        &\leq
                         e^*_{m},
\end{align*}
so~$e^*_{m-2}\leq 1/2$.
Proceeding in this way yields
$ \lambda_m^* \leq \lambda_{m-1}^*  \leq \dots \leq \lambda_{1}^* \leq \lambda_{0}^*$. This contradicts~\eqref{eq:l0l1_ratio}, and thus  proves that
\[
          \lambda_{m-1}^* < \lambda_m^*  .
\]
We can now prove, in a similar fashion, that~$ \lambda_{m-2}^* < \lambda_{m-1}^*$ and then that
$\lambda_{m-3}^* < \lambda_{m-2}^*$, etc.,
and this yields~\eqref{eq:ntsq}. Applying Prop.~\ref{prop:lambda_ratio} yields~$e^*_i>1/2$,
$i=1,\dots,m$. Combining this with~\eqref{eq:ntsq},~\eqref{eq:list},
and~\eqref{eq:eismono} proves~\eqref{eq:niseven}.

We now turn to the  case where~$n$ is odd. Let~$m:=(n-1)/2$.
Note that~\eqref{eq:eismono} implies that~$e^*_{m+1}=1/2$.
We need to show that~\eqref{eq:ntsq} holds.
 Seeking a contradiction, assume that
\be\label{eq:l2l1_odd}
\lambda_{m-1}^* \geq \lambda_m^*.
\ee
Then~\eqref{eq:lambda_ratio} gives~$e_m^*\leq 1/2$. Now~\eqref{eq:ep} and~\eqref{eq:eisnodd}
 yield
\begin{align*}
                        e^*_{m-1}&=\frac{\lambda^*_m}{\lambda^*_{m-1}}
                        \frac{e^*_m}{1-e^*_m}(1-e^*_{m+1})\\
                        &\leq 1\times 1\times 1/2.
\end{align*}
  Combining this with~\eqref{eq:lambda_ratio} yields
$ \lambda_{m-2}^* \geq \lambda_{m-1}^* $.
 Proceeding as above yields
$ \lambda_m^* \leq \lambda_{m-1}^* \leq  \dots \leq \lambda_{1}^* \leq \lambda_{0}^*$. This contradicts~\eqref{eq:l0l1_ratio}, and thus  proves that
\[
          \lambda_{m-1}^* < \lambda_m^*  .
\]
We can now prove, in a similar fashion, that~$ \lambda_{m-2}^* < \lambda_{m-1}^*$ and then that
$\lambda_{m-3}^* < \lambda_{m-2}^*$, etc.,
and this yields~\eqref{eq:ntsq}. The proof
of~\eqref{eq:eisnodd} follows
as in the case where~$n$ is even. This completes the proof of Prop.~\ref{prop:lambda_ratio_hom}.~\IEEEQED

\section*{Acknowledgment}
 We thank   Leonid Mirkin for helpful discussions. This research is partially supported by research grants
from  the  ISF and from the Ela Kodesz Institute for Medical Engineering and Physical Sciences.

\bibliographystyle{IEEEtranS}

\bibliography{RFM_bibl_for_rfm_sense,RFM_bibl}

\end{document}